# Magnetic anisotropy driven by ligand in 4*d* transition-metal oxide SrRuO$_3$

Yuki K. Wakabayashi,[1,*] Masaki Kobayashi,[2,3] Yuichi Seki,[3] Yoshinori Kotani,[4] Takuo Ohkochi,[4] Kohei Yamagami,[4] Miho Kitamura,[5] Yoshitaka Taniyasu,[1] Yoshiharu Krockenberger,[1] and Hideki Yamamoto[1]

[1]*NTT Basic Research Laboratories, NTT Corporation, Atsugi, Kanagawa 243-0198, Japan*
[2]*Center for Spintronics Research Network, The University of Tokyo, 7-3-1 Hongo, Bunkyo-ku, Tokyo 113-8656, Japan*
[3]*Department of Electrical Engineering and Information Systems, The University of Tokyo, Bunkyo, Tokyo 113-8656, Japan*
[4]*Japan Synchrotron Radiation Research Institute (JASRI), 1-1-1 Kouto, Sayo, Hyogo, 679-5198, Japan*
[5]*Institute for Advanced Synchrotron Light Source, National Institutes for Quantum Science and Technology (QST), Sendai 980-8579, Japan*

[*]Corresponding author: yuuki.wakabayashi@ntt.com

Abstract

The origin of magnetic anisotropy in magnetic compounds is a longstanding issue in solid state physics and nonmagnetic ligand ions are considered to contribute little to magnetic anisotropy. Here, we introduce the concept of ligand-driven magnetic anisotropy in a complex transition-metal oxide. We conducted X-ray absorption and X-ray magnetic circular dichroism spectroscopies at the Ru and O edges in the 4*d* ferromagnetic metal SrRuO$_3$. Systematic variation of the sample thickness in the range of ≤ 10 nm allowed us to control the localization of Ru 4*d* $t_{2g}$ states, which affects the magnetic coupling between the Ru and O ions. We found that the orbital magnetization of the ligand induced via hybridization with the Ru 4*d* orbital determines the magnetic anisotropy in SrRuO$_3$.

Magnetic anisotropy that confers a preferred direction on the spin of a system is one of the most important properties of magnetic materials. Over the long history of studies on magnetic materials, various origins have been suggested for magnetic anisotropy: magnetic multipoles [1,2], electric quadrupole [3-6], and single-ion anisotropy, where the focus has been on crystal field splitting at $d$- and/or $f$- orbitals of transition metal elements [7-9]. Especially in magnetic oxides, such as perovskite manganites [6,10], ruthenates [11-13], spinel ferrites [14], pyrochlores [15], and their epitaxial heterostructures [16], the electric quadrupole moment due to the different $d$ orbital occupations and the anisotropy of the orbital magnetic moment based on the single-ion model or tight-binding model [17] have been considered to be the two main origins of magnetic anisotropy. Elucidating the origin of magnetic anisotropy in such magnetic oxides is vital for comprehending magnetism as well as utilizing spin degrees of freedom in oxide electronics [18,19].

The itinerant 4$d$ ferromagnetic perovskite ruthenate SrRuO$_3$ (SRO) [bulk Curie temperature ($T_C$) = 165 K] has been used as a model system for understanding magnetic anisotropy in metallic oxides [20-29]. In addition, recent observations of Weyl fermions in SRO have brought renewed interest as an emergent platform for exploring quantum transport phenomena for topological oxide electronics [30-34]. In SRO films, the magnetic anisotropy is linked to epitaxial strain through the interplay between electrons (charges, spins), lattice, and orbital degrees of freedom [35-39]. Particularly, compressively strained SRO film on SrTiO$_3$ (STO) (001) was the first oxide heterostructure in which perpendicular magnetic anisotropy was discovered [40]. The perpendicular magnetic anisotropy in compressively strained SRO has been considered to arise from the anisotropy of the Ru orbital magnetic moment [11-13]. However, our recent observations of the isotropic Ru orbital magnetic moment relative to the spin magnetic moment $m_{orb}/m_{spin}$ in compressively strained SRO films contradict the conventional understanding [41]; the scenario in which magnetic anisotropy originates from an anisotropic variation in the crystal field splitting at the 4$d$ orbitals of Ru due to strain [10-12] does not hold true. This implies that contributions from the ligand O 2$p$ orbital should also be considered. In fact, polarized neutron scattering experiments have shown that O ions have unexpectedly large magnetic moments; the contribution to the total magnetization reaches 30% [42]. The contribution from oxygen may come from hybridization between Ru 4$d$ $t_{2g}$ and O 2$p$ orbitals in the itinerant system SRO. Since an increase in localization of the Ru 4$d$ $t_{2g}$ states shrinks the wave function, resulting in lower hybridization strength, experiments that systematically vary the degree of localization of Ru 4$d$ $t_{2g}$ states would provide cogent evidence for the oxygen ligand-driven magnetic anisotropy. A possible approach to controlling the degree of localization of Ru 4$d$ $t_{2g}$ states is the introduction of interface-driven defects by controlling film thickness or an increase in the effective Coulomb interaction with dimensionality reduction [43]. The recent advances in machine-learning-assisted thin-film growth technology [44,45] have enabled systematic and precise control of the thickness of stoichiometric SRO films in the range of $\leq$ 10 nm.

In this study, we investigated the anisotropy of the Ru 4$d$ and O 2$p$ electronic structure and the magnetic properties of stoichiometric compressive-strained SRO films of various thicknesses $t$ = 1-10 nm grown on STO (001) substrates by using soft X-ray absorption spectroscopy (XAS) and X-ray magnetic circular dichroism (XMCD), which are highly sensitive to the local electronic structure and element-specific magnetic properties in magnetic materials [46-49]. By varying the localization of the Ru 4$d$ $t_{2g}$ states through thickness control, we found that the perpendicular magnetic anisotropy originates from the anisotropy of the magnetic moments of O 2$p$ orbitals. This finding is conceptually distinct from the conventional single-ion model that is based on the effect of the crystal field at the $d$ orbitals of magnetic elements.

We grew compressively strained epitaxial SRO films with a thickness $t$ of 1, 2, 3, 4, 5, and 10 nm on STO (001) substrates [Fig. 1(a)] in a custom-designed molecular beam epitaxy (MBE) system [44]. The residual resistivity ratio (RRR) of the thickest SRO film ($t$ = 10 nm) was 21. This is high enough to observe the intrinsic quantum transport of Weyl fermions in SRO [50] and is a hallmark of the stoichiometric composition of the films [51]. The XAS and XMCD measurements were performed at the helical undulator beamline BL25SU of SPring-8 [52]. For the XMCD measurements, absorption spectra for circularly polarized X rays with the photon helicity parallel ($\mu^+$) or antiparallel ($\mu^-$) to the spin polarization were obtained by reversing the photon helicity at each photon energy $h\nu$ and recorded in the total-electron-yield (TEY) mode. The measurement temperature was 20 K, which was low enough considering the $T_C$ of the SRO films (100-150 K) [50]. All XAS and XMCD spectroscopy measurements were performed at 1.92 T, the maximum applied magnetic field in the system [53]. To measure the perpendicular and in-plane magnetic moments, the angles of the external magnetic fields and incident X rays measured from the sample surfaces were set to $\theta$ = 90° and 20°, respectively [Figs. 1(b) and 1(c)]. The incident light was tilted at 10° to the magnetic field [53]. The details of the samples, growth conditions, and experimental set up for the XAS and XMCD measurements are described in Sec. I of the Supplemental Material [54].

We will start with the XAS and XMCD spectra of the thickest film (10 nm) taken in the perpendicular configuration ($\theta$ = 90°). Figure 2(a) shows the Ru $M_{2,3}$-edge XAS and XMCD spectra for the SRO film with $t$ = 10 nm. The spectral features are essentially identical to what we have reported for much thicker (60-63 nm) SRO films on STO [13, 41]. The absorption peaks from the Ru $3p_{3/2}$ and $3p_{1/2}$ core levels into the Ru 4$d$ states appear at around 463 and 485 eV, respectively, both in the XAS and XMCD spectra [55,56]. The details of the peak assignments of the Ru $M_{2,3}$-edge XAS and XMCD spectra are in Sec. II of the Supplemental Material [54]. To further investigate the unoccupied electronic states hybridized with the O 2$p$ orbitals and the magnetic properties of the O ions, we also measured the O $K$-edge XAS and XMCD spectra of the 10-nm SRO film under the same conditions [Fig. 2(b)]. The absorption peak at 529 eV arises from transitions to the Ru 4$d$ $t_{2g}$ states hybridized with the O 2$p$ states, while transitions to the Ru 4$d$ $e_g$ states are observed in the energy range of 531.5–534.2 eV [57,58]. The range of 534.5–539.5 eV displays transitions to the Sr 4$d$ states, and transitions to the Ru 5$s$ states occur in 540–546 eV [57,58]. The intense coherent Ru 4$d$ $t_{2g}$ peak [41] indicates the long lifetimes of quasiparticles in the hybridized Ru 4$d$ $t_{2g}$-O 2$p$ states, which is also evidenced by observation of quantum oscillations in the resistivity of SRO film with $t$ = 10 nm [50]. A substantial orbital magnetic moment of the O 2$p$ states was deduced from the O $K$ XMCD spectrum in the energy range corresponding to Ru 4$d$ $t_{2g}$ [Fig. 2(b)], indicating that spin-polarized bands crossing the Fermi level, $E_F$, are formed by the Ru 4$d$ $t_{2g}$ states hybridized with the O 2p states and that there was charge transfer [59] from the ligand O 2p orbitals to the Ru 4d orbitals. This interpretation is in accordance with our previous density functional theory (DFT) calculations [31], which show that the half-metallic hybridized Ru 4d $t_{2g}$-O 2p states cross $E_F$.

Next, we turn to the $t$ and $\theta$ dependences. When the film thickness was reduced to half ($t$ = 5 nm), similar Ru $M_{2,3}$- and O $K$-edge spectra to those above were observed [Figs. S1(a) and S1(e) of the Supplemental Material [54]]. Upon decreasing the film thickness ($t \leq 4$ nm) further, absorptions from the STO substrate [60] became prominent [Figs. 2(c) and 2(d) for $t$ = 1 nm, Figs. S1(b)-S1(d) and S1(f)-S1(h) of the Supplemental Material [54] for $t$ = 2-4 nm]. As in the Ru $M_{2,3}$-edge XAS spectrum for $t$ = 1 nm, the Ti $L_{2,3}$-edge absorptions were convoluted over the Ru $M_{2,3}$ edges [Fig. 2(c)]. Similarly, the O $K$-edge XAS spectra for $t$ = 1 nm showed absorptions from the STO substrate, such as the absorption peak at 530.7 eV arising from transitions from O 1$s$ states to the Ti 3$d$ $t_{2g}$ states [Fig. 2(d)] [60]. Nonetheless, the Ru $M_{2,3}$-edge and O $K$-edge XMCD could selectively extract information on the SRO films, since the Ti ions in the STO substrates are nonmagnetic. The Ru $M_{2,3}$-edge XMCD spectra normalized at the Ru $M_2$-edge XMCD peak taken at $\theta$ = 90° are plotted in Fig. 3(a), in order to see the evolution of the $m_{orb}/m_{spin}$ ratio with $t$. From the XMCD sum rules, $m_{orb}/m_{spin}$ ratio is determined solely from the ratio of the $M_3$-edge to $M_2$-edge XMCD absorption intensities [see Sec. IV of the Supplemental Material [54]]. The spectra almost completely overlap, indicating that the Ru$^{4+}$ 4$d$ states have a common $m_{orb}/m_{spin}$ ratio (0.081), as estimated from the XMCD sum rules [46,47,48], irrespective of $t$. This is in contrast to the case of SRO thin films prepared by pulsed-laser deposition (PLD), where ultrathin (1-2 nm) films became paramagnetic and showed a significant change in electronic structure [12]. This discrepancy may have arisen from the inevitable formation of Ru vacancies in the PLD-grown films [61]. We also compared the normalized Ru $M_{2,3}$-edge XMCD spectra for different $\theta$ (90° and 20°) and found that they were identical, as exemplified by the data for the $t$ = 2 nm film in Fig. 3(a). Altogether, the $t$ and $\theta$ dependencies of XMCD indicate that the electronic structure and orbital magnetic moment are isotropic within the Ru$^{4+}$ states.

Now that the magnetic anisotropy in SRO tuned out not to arise from the supposed anisotropy in the orbital magnetic moment of Ru, the key to clarifying the origin of the magnetic anisotropy becomes estimating the individual magnetic moments of Ru and O themselves. In XMCD measurements, the magnetic moments of each element are proportional to the ratios of XMCD to XAS intensities of the particular absorption peaks. However, the presence of absorptions from the STO substrates in the XMCD spectra for the SRO films with $t$ = 1-4 nm prevented a quantitative evaluation of the variation in the magnetic moment with film thickness and direction of the applied magnetic field. To overcome this difficulty, we developed an analytical procedure to subtract the signals from the substrate. The details of the deconvolution procedure and examples of using it on the XAS spectra of SRO film are described in Sec. V of the Supplemental Material [54].

Using the O $K$-edge XAS spectra inherent to each thickness, we revealed that the intensity of the coherent Ru 4$d$ $t_{2g}$ peak decreases with decreasing $t$ [Fig. 3(b)], indicating that the lifetimes of the quasiparticles in the hybridized Ru $4d$ $t_{2g}$-O $2p$ states become shorter and the localization of the Ru $4d$ $t_{2g}$ states become stronger due to disorder-induced localization near the interface or an increase in the effective Coulomb interaction with dimensionality reduction [62,63]. Here, the O $K$-edge XAS spectra were normalized at the energy, where coherent absorption is reduced and incoherent absorption begins. The increase in the localization of the Ru $4d$ $t_{2g}$ states shrinks the wave function, resulting in weaker hybridization. The reduction in peak intensity became prominent when $t$ was reduced from 2 nm to 1 nm, which is consistent with the change in the global transport properties [50]: metallicity was barely observed in the SRO film with $t$ = 1nm, while localization was noticeable at 20 K, the measurement temperature of XAS in this study.

Figure 3(c) shows the thickness dependence of the ratios of the XMCD intensity to the XAS intensity of the Ru $M_2$ and O $K$-edge Ru $4d$ $t_{2g}$ peaks at $\theta$ = 90°, which are proportional to the magnetic moments of Ru and O ions, respectively. Note that XMCD at the $K$ edges is only able to probe the orbital magnetic moment [46]. The orbital magnetic moment of oxygen (red circles) decreases with decreasing $t$. This trend coincides well with that of the O $K$-edge Ru $4d$ $t_{2g}$ peak intensity (blue triangles), suggesting that the magnetic moment of oxygen decreases as a result of reduced charge transfer through orbital hybridization. In addition, the magnetic moment of Ru (green circles) decreases significantly below 2 nm, indicating that the Ru $4d$ $t_{2g}$-O $2p$ hybridization is also important for the ferromagnetic ordering of Ru. Note that the conclusion of weakened ferromagnetic ordering in SRO films with $t \leq 2$ nm is supported by the thickness-dependent Curie temperature ($T_C$) determined from transport measurements [51]. To investigate the change in magnetic anisotropy of Ru and O ions with respect to the Ru $4d$ $t_{2g}$-O $2p$ hybridization, we compared the ratios of the XMCD intensity to the XAS intensity in the perpendicular configuration ($\theta$ = 90°) and in the in-plane configuration ($\theta$ = 20°) [Fig. 3(d)]. All the XAS and XMCD spectra with $t$ = 1-10 nm at $\theta$ = 20° for extracting the data points in Fig. 3(d) are in Fig. S2 of the Supplemental Material [54]. The ratio represents the in-plane/perpendicular magnetization ratio $r$, which means the magnetic moment has perpendicular magnetic anisotropy when $r$ is less than 1. For $t \geq 3$ nm, except for $t$ = 5 nm, $r$ was 0.4-0.6 for Ru and O. Only $t$ = 5 nm showed a slightly weaker perpendicular magnetic anisotropy, with $r$ being 0.7-0.8 for Ru and O. Although there were variations in the magnitude of the magnetic anisotropy from sample to sample, the $r$ values for Ru and O were almost the same at each thickness, indicating that Ru and O had the same magnetic anisotropy and were magnetically coupled. Notably, the magnetic anisotropies of Ru and O were different in the SRO films with $t$ = 1 and 2 nm; O had an $r$ of about 0.7, while Ru had an $r$ of about 1.0, i.e., no magnetic anisotropy. As described earlier, the orbital hybridization became stronger in the range $t \geq 3$ nm. The perpendicular magnetic anisotropy of Ru for $t \geq 3$ nm is considered to be due to the magnetic coupling between O and Ru through orbital hybridization. In other words, the perpendicular magnetic anisotropy of SRO is due to not the Ru orbitals but rather the O orbitals and is induced through the Ru $4d$ $t_{2g}$-O $2p$ orbital hybridization. Changes in the magnetic properties of Ru for the SRO film with $t$ = 1 and 2 nm were also found by studying the Ru $M_3$-edge XMCD-$\mu_0 H$ curves [see Sec. VI of the Supplemental Material [54]].

Conventionally, the magnetic anisotropy in SRO films has been interpreted in terms of a strain-induced anisotropic variation in the crystal field splitting at the $d$ orbitals of Ru (single-ion picture) [10-12]. Such an interpretation contradicts the new finding in this study that the magnetic anisotropy is driven by oxygen ligands. The magnetic moment of $Ru^{4+}$ ions also shows the breakdown of the single-ion model, as follows. In the case of a large crystal field splitting, $Ru^{4+}$ ions are expected to be in a low-spin state with an electron configuration of $t_{2g}^4$ (3↑, 1↓) corresponding to 2 $\mu_B$/$Ru^{4+}$ ions [11,23]. However, the total magnetic moment, $m_{total} = m_{spin} + m_{orb}$, estimated from the XAS and XMCD spectra in Fig. 2(a) by using the XMCD sum rules [46-48] is 0.71 $\mu_B$/$Ru^{4+}$, indicating a limitation to the local picture. Recently, polarized neutron diffraction measurements on bulk SRO were used to observe the cubic faces of the Ru spin density distribution and the disk-shaped O spin density distribution, both perpendicular to the Ru $4d$ $t_{2g}$-O $2p$ bonds[42]; the results suggested that, for the Ru-O bond along the $x$ direction, the $d_{xy}$ and $d_{xz}$ form π bonds with oxygen $p_y$ and $p_z$ orbitals, respectively. A possible origin of the ligand-driven magnetic anisotropy found in this study is a strain-induced change in such anisotropic spin distributions at the O ions. It has been widely acknowledged that Ru $4d$ $t_{2g}$-O $2p$ hybridization, as is evidenced by the intense coherent O $K$-edge Ru $4d$ $t_{2g}$ peak [Fig. 4(a)], is the key to ferromagnetism and metallicity emerging in SRO [31,64]. The present study indicates that it is also the case for magnetic anisotropy.

Recently, magnetic moments of nonmagnetic ligand induced by charge transfer through the orbital hybridization from $3d$ or $5d$ orbitals were measured in the $3d$ and $5d$ ferromagnetic compounds [65,66], confirming the significant contribution of ligands to the ferromagnetism in $3d$ and $5d$ systems. Exploring the existence of ligand-driven magnetic anisotropy even in such $3d$ and $5d$ systems would be crucial for universal understanding of the origin of magnetic anisotropy. In systems where magnetic moments are induced in nonmagnetic ligands through $p$-$d$ hybridization, ligand-driven magnetic anisotropy could dominate the magnetic anisotropy.

In conclusion, a new type of magnetic anisotropy, ligand-driven magnetic anisotropy, has been discovered in $4d$ ferromagnetic Weyl semimetal SRO by making XAS and XMCD measurements. Systematic variation of the film thickness in the range of ≤ 10 nm for controlling the localization of the Ru $4d$ $t_{2g}$ states revealed that the magnetic moments of the nonmagnetic ligand (O) are induced by charge transfer through the orbital hybridization from Ru $4d$ orbitals, and the anisotropy of the ligand determines the magnetic anisotropy of SRO. The ligand-driven magnetic anisotropy is distinct from the conventional single-ion model dominated by the effects of the crystal field splitting at the $d$ orbitals of magnetic elements. Since this new concept can be extended to other magnetic materials, further theoretical calculations and experimental investigations on element-specific electronic structures near heterointerfaces are encouraged in order to elucidate the mechanism behind the emergence of anisotropy in the ligand.

**ACKNOWLEDGMENTS**

This work was supported by a Grants-in-Aid (Grants No. 22H04948), of the Japan Science and Technology Agency. This work was partially supported by the Spintronics Research Network of Japan (Spin-RNJ). The synchrotron radiation experiments were performed at the BL25SU of SPring-8 with the approval of the Japan Synchrotron Radiation Research Institute (JASRI) (Proposals No. 2022B1403 and No. 2023A1369).

**DATA AVAILABILITY**
The data that support the findings of this study are available from the corresponding authors upon reasonable request.

**References**
[1] T. Matsumura, T. Yonemura, K. Kunimori, M. Sera, and F. Iga, Magnetic Field Induced $4f$ Octupole in $CeB_6$ Probed by Resonant X-Ray Diffraction, Phys. Rev. Lett. **103**, 017203 (2009).
[2] Y. Shimizu, A. Miyake, A. Maurya, F. Honda, A. Nakamura, Y. J. Sato, D. Li, Y. Homma, M. Yokoyama, Y. Tokunaga, M. Tokunaga, and D. Aoki, Strong magnetic anisotropy and unusual magnetic field reinforced phase in URhSn with a quasi-kagome structure, Phys. Rev. B **102**, 134411 (2020).
[3] D. S. Wang, R. Wu, and A. J. Freeman, First-principles theory of surface magnetocrystalline anisotropy and the diatomic-pair model. Phys. Rev. B **47**, 14932 (1993).
[4] G. van der Laan, Microscopic origin of magnetocrystalline anisotropy in transition metal thin films. J. Phys. Condens. Matter. **10**, 3239 (1998).
[5] J. Okabayashi, Y. Miura, Y. Kota, K. Z. Suzuki, A. Sakuma, and S. Mizukami, Detecting quadrupole: a hidden source of magnetic anisotropy for Manganese alloys, Sci. Rep. **10**, 9744 (2020).
[6] G. Shibata, M. Kitamura, M. Minohara, K. Yoshimatsu, T. Kadono, K. Ishigami, T. Harano, Y. Takahashi, S. Sakamoto, Y. Nonaka, K. Ikeda, Z. Chi, M. Furuse, S. Fuchino, M. Okano, J.I. Fujihira, A. Uchida, K. Watanabe, H. Fujihira, S. Fujihira, A. Tanaka, H. Kumigashira, T. Koide, and A. Fujimori, Anisotropic spin-density distribution and magnetic anisotropy of strained $La_{1-x}Sr_xMnO_3$ thin films: Angle-dependent X-ray magnetic circular dichroism, npj Quantum Mater. **3**, 3 (2018).
[7] J. C. Slonczewski, Anisotropy and Magnetostriction in Magnetic Oxides, J. Appl. Phys. 32, S253 (1961).
[8] G. A. Craig and M. Murrie, 3d single-ion magnets, Chem. Soc. Rev. 44, 2135 (2015).
[9] C. H. Sohn, C. H. Kim, L. J. Sandilands, N. T. M. Hien, S. Y. Kim, H. J. Park, K. W. Kim, S. J. Moon, J. Yamaura, Z. Hiroi, and T. W. Noh, Strong Spin-Phonon Coupling Mediated by Single Ion Anisotropy in the All-In-All-Out Pyrochlore Magnet $Cd_2Os_2$

O$_7$, Phys. Rev. Lett. **118**, 117201 (2017).

[10] M. Kobayashi, L. D. Anh, M. Suzuki, S. Kaneta-Takada, Y. Takeda, S. I. Fujimori, G. Shibata, A. Tanaka, M. Tanaka, S. Ohya, and A. Fujimori, Alternation of Magnetic Anisotropy Accompanied by Metal-Insulator Transition in Strained Ultrathin Manganite Heterostructures, Phys. Rev. Appl. **15**, 064019 (2021).

[11] A. J. Grutter, F. J. Wong, E. Arenholz, A. Vailionis, and Y. Suzuki, Evidence of high-spin Ru and universalmagnetic anisotropy in SrRuO$_3$ thin films, Phys. Rev. B **85**, 134429 (2012).

[12] K. Ishigami, K. Yoshimatsu, D. Toyota, M. Takizawa, T. Yoshida, G. Shibata, T. Harano, Y. Takahashi, T. Kadono, V. K. Verma, V. R. Singh, Y. Takeda, T. Okane, Y. Saitoh, H. Yamagami, T. Koide, M. Oshima, H. Kumigashira, and A. Fujimori, Thickness-dependent magnetic properties and strain-induced orbital magnetic moment in SrRuO$_3$ thin film, Phys. Rev. B **92**, 064402 (2015).

[13] Y. K. Wakabayashi, M. Kobayashi, Y. Takeda, K. Takiguchi, H. Irie, S.-i. Fujimori, T. Takeda, R. Okano, Y. Krockenberger, Y. Taniyasu, and H. Yamamoto, Single-domain perpendicular magnetization induced by the coherent O 2$p$-Ru 4$d$ hybridized state in an ultra-high-quality SrRuO$_3$ film Phys. Rev. Mater., **5** 124403 (2021).

[14] J. Okabayashi, M. A. Tanaka, M. Morishita, H. Yanagihara, and K. Mibu, Origin of perpendicular magnetic anisotropy in Co$_x$Fe$_{3-x}$O$_{4+\delta}$ thin films studied by x-ray magnetic circular and linear dichroisms, Phys. Rev. B **105**, 134416 (2022).

[15] H. Shinaoka, T. Miyake, and S. Ishibashi, Noncollinear Magnetism and Spin-Orbit Coupling in 5$d$ Pyrochlore Oxide Cd$_2$Os$_2$O$_7$, Phys. Rev. Lett. **108**, 247204 (2012).

[16] Z. Xiao, F. Zhang, M. A. Farrukh, R. Wang, G. Zhou, Z. Quan, and X. Xu, Perpendicular magnetic anisotropy in compressive strained La$_{0.67}$Sr$_{0.33}$MnO$_3$ films, J. Mater. Sci. **54**, 9017 (2019).

[17] P. Bruno, Tight-binding approach to the orbital magnetic moment and magnetocrystalline anisotropy of transition-metal monolayers, Phys. Rev. B **39**, 865 (1989).

[18] K. S. Takahashi, A. Sawa, Y. Ishii, H. Akoh, M. Kawasaki, and Y. Tokura, Inverse tunnel magnetoresistance in all-perovskite junctions of La$_{0.7}$Sr$_{0.3}$MnO$_3$/SrTiO$_3$/SrRuO$_3$, Phys. Rev. B **67**, 094413 (2003).

[19] L. Liu, Q. Qin, W. Lin, C. Li, Q. Xie, S. He, X. Shu, C. Zhou, Z. Lim, J. Yu, W. Lu, M. Li, X. Yan, S. J. Pennycook, and J. Chen, Current-induced magnetization switching in all-oxide heterostructures, Nat. Nanotechnol. **14**, 939 (2019).

[20] J. J. Randall and R. Ward, The Preparation of Some Ternary Oxides of the Platinum Metals, J. Am. Chem. Soc. **81**, 2629 (1959).

[21] L. Klein, J. S. Dodge, C. H. Ahn, J. W. Reiner, L. Mieville, T. H. Geballe, M. R. Beasley, and A. Kapitulnik, Transport and magnetization in the badly metallic itinerant ferromagnet SrRuO$_3$, J. Phys. Condens. Matter **8**, 10111 (1996).

[22] C. B. Eom, R. J. Cava, R. M. Fleming, J. M. Phillips, R. B. van Dover, J. H. Marshall, J. W. P. Hsu, J. J. Krajewski, and W. F. Peck, Jr., Single-Crystal Epitaxial Thin Films

of the Isotropic Metallic Oxides Sr$_{1-x}$Ca$_x$RuO$_3$ (0 ≤ $x$ ≤ 1), Science **258**, 1766, (1992).

[23] G. Koster, L. Klein, W. Siemons, G. Rijnders, J. S. Dodge, C.-B. Eom, D. H. A. Blank, M. R. Beasley, Structure, physical properties, and applications of SrRuO$_3$ thin films, Rev. Mod. Phys. **84**, 253 (2012).

[24] Y. K. Wakabayashi, Y. Krockenberger, T. Otsuka, H. Sawada, Y. Taniyasu, and H. Yamamoto, Intrinsic physics in magnetic Weyl semimetal SrRuO$_3$ films addressed by machine-learning-assisted molecular beam epitaxy, Jpn. J. Appl. Phys. **62**, SA0801 (2023).

[25] D. Kan, M. Mizumaki, T. Nishimura, and Y. Shimakawa, Orbital Magnetic Moments in Strained SrRuO$_3$ Thin Films, Phys. Rev. B, **94**, 214420, (2016).

[26] D. E. Shai, C. Adamo, D. W. Shen, C. M. Brooks, J. W. Harter, E. J. Monkman, B. Burganov, D. G. Schlom, and K.M. Shen, Quasiparticle mass enhancement and temperature dependence of the electronic structure of ferromagnetic SrRuO$_3$ thin films, Phys. Rev. Lett. **110**, 087004 (2013).

[27] A. P. Mackenzie, J. W. Reiner, A. W. Tyler, L. M. Galvin, S. R. Julian, M. R. Beasley, T. H. Geballe, and A. Kapitulnik, Observation of quantum oscillations in the electrical resistivity of SrRuO$_3$, Phys. Rev. B **58**, R13318 (1998).

[28] Z. Li, S. Shen, Z. Tian, K. Hwangbo, M. Wang, Y. Wang, F. M. Bartram, L. He, Y. Lyu, Y. Dong, G. Wan, H. Li, N. Lu, J. Zang, H. Zhou, E. Arenholz, Q. He, L. Yang, W. Luo, and P. Yu, Reversible manipulation of the magnetic state in SrRuO$_3$ through electric-field controlled proton evolution, Nat. Commun. **11**, 184 (2020).

[29] H. Boschker, T. Harada, T. Asaba, R. Ashoori, A. V. Boris, H. Hilgenkamp, C. R. Hughes, M. E. Holtz, L. Li, D. A. Muller, H. Nair, P. Reith, X. R. Wang, D. G. Schlom, A. Soukiassian, and J. Mannhart, Ferromagnetism and Conductivity in Atomically Thin SrRuO$_3$, Phys. Rev. X **9**, 011027 (2019).

[30] Y. Chen, D. L. Bergman, and A. A. Burkov, Weyl fermions and the anomalous Hall effect in metallic ferromagnets, Phys. Rev. B **88**, 125110 (2013).

[31] K. Takiguchi, Y. K. Wakabayashi, H. Irie, Y. Krockenberger, T. Otsuka, H. Sawada, S. A. Nikolaev, H. Das, M. Tanaka, Y. Taniyasu, and H. Yamamoto, Quantum transport evidence of Weyl fermions in an epitaxial ferromagnetic oxide, Nat. Commun. **11**, 4969 (2020).

[32] S. K. Takada, Y. K. Wakabayashi, Y. Krockenberger, T. Nomura, Y. Kohama, H. Irie, K. Takiguchi, S. Ohya, M. Tanaka, Y. Taniyasu, and H. Yamamoto, High-mobility two-dimensional carriers from surface Fermi arcs in magnetic Weyl semimetal films, npj Quantum Mater. **7**, 102 (2022).

[33] U. Kar, A. K. Singh, Y.-T. Hsu, C.-Y. Lin, B. Das, C.-T. Cheng, M. Berben, S. Yang, C.-Y. Lin, C.-H. Hsu, S. Wiedmann, W.-C. Lee, and W.-L. Lee, The thickness dependence of quantum oscillations in ferromagnetic Weyl metal SrRuO$_3$, npj Quantum Mater. **8**, 8 (2023).

[34] W. Lin, L. Liu, Q. Liu, L. Li, X. Shu, C. Li, Q. Xie, P. Jiang, X. Zheng, R. Guo, Z. Lim, S. Zeng, G. Zhou, H. Wang, J. Zhou, P. Yang, Ariando, S. J. Pennycook, X. Xu,

Z. Zhong, Z. Wang, and J. Chen, Electric Field Control of the Magnetic Weyl Fermion in an Epitaxial SrRuO$_3$ (111) Thin Film, Adv. Mater. **3**, 2101316 (2021).

[35] Q. Gan, R. A. Rao, C. B. Eom, J. L. Garrett, and M. Lee, Direct measurement of strain effects on magnetic and electrical properties of epitaxial SrRuO$_3$ thin films. Appl. Phys. Lett. **72**, 978 (1998).

[36] C. U. Jung, H. Yamada, M. Kawasaki, Y. Tokura, Magnetic anisotropy control of SrRuO$_3$ films by tunable epitaxial strain, Appl. Phys. Lett. **84**, 2590 (2004).

[37] A. T. Zayak, X. Huang, J. B. Neaton, and K. M. Rabe, Structural, electronic, and magnetic properties of SrRuO$_3$ under epitaxial strain, Phys. Rev. B **74**, 094104 (2006).

[38] D. Kan, R. Aso, H. Kurata, and Y. Shimakawa, Epitaxial strain effect in tetragonal SrRuO$_3$ thin films, J. Appl. Phys. **113**, 173912 (2013).

[39] Y. K. Wakabayashi, S. K. Takada, Y. Krockenberger, Y. Taniyasu, and H. Yamamoto, Wide-range epitaxial strain control of electrical and magnetic properties in high-quality SrRuO$_3$ films, ACS Applied Electronic Materials **3**, 2712 (2021).

[40] L. Klein, J. S. Dodge, T. H. Geballe, A. Kapitulnik, A. F. Marshall, L. Antognazza, and K. Char, Perpendicular magnetic anisotropy and strong magneto-optic properties of SrRuO$_3$ epitaxial films, Appl. Phys. Lett. **66**, 2427 (1995).

[41] Y. K. Wakabayashi, M. Kobayashi, Y. Takeda, M. Kitamura, T. Takeda, R. Okano, Y. Krockenberger, Y. Taniyasu, and H. Yamamoto, Isotropic orbital magnetic moments in magnetically anisotropic SrRuO$_3$ Films, Phys. Rev. Mater. 6, 094402 (2022).

[42] S. Kunkemöller, K. Jenni, D. Gorkov, A. Stunault, S. Streltsov, and M. Braden, Magnetization density distribution in the metallic ferromagnet SrRuO$_3$ determined by polarized neutron diffraction, Phys. Rev. B **100**, 054413 (2019).

[43] H. Jeong, S. G. Jeong, A. Y. Mohamed, M. Lee, W. Noh, Y. Kim, J. S. Bae, W. S. Choi, and D. Y. Cho, Thickness-dependent orbital hybridization in ultrathin SrRuO$_3$ epitaxial films, Appl. Phys. Lett. **115**, 092906 (2019).

[44] Y. K. Wakabayashi, T. Otsuka, Y. Krockenberger, H. Sawada, Y. Taniyasu, and H. Yamamoto, Machine-learning-assisted thin-film growth: Bayesian optimization in molecular beam epitaxy of SrRuO$_3$ thin films, APL Mater. **7**, 101114 (2019).

[45] Y. K. Wakabayashi, T. Otsuka, Y. Krockenberger, H. Sawada, Y. Taniyasu, H. Yamamoto, Bayesian optimization with experimental failure for high-throughput materials growth, npj Comput. Mater. **8**, 180 (2022).

[46] B. T. Thole, P. Carra, F. Sette, and G. van der Laan, X-ray circular dichroism as a probe of orbital magnetization, Phys. Rev. Lett. **68**, 1943 (1992).

[47] C. T. Chen, Y. U. Idzerda, H. -J. Lin, N. V. Smith, G. Meigs, E. Chaban, G. H. Ho, E. Pellegrin, and F. Sette, Experimental confirmation of the X-ray magnetic circular dichroism sum rules for iron and cobalt, Phys. Rev. Lett. **75**, 152 (1995).

[48] J. Stohr and H. Konig, Determination of spin-and orbital-moment anisotropies in transition metals by angle-dependent X-ray magnetic circular dichroism, Phys. Rev. Lett. **75**, 3748 (1995).

[49] Y. K. Wakabayashi, Y. Nonaka, Y. Takeda, S. Sakamoto, K. Ikeda, Z. Chi, G. Shibata,

A. Tanaka, Y. Saitoh, H. Yamagami, M. Tanaka, A. Fujimori, and R. Nakane, Electronic structure and magnetic properties of magnetically dead layers in epitaxial $CoFe_2O_4/Al_2O_3/Si$ (111) films studied by x-ray magnetic circular dichroism, Phys. Rev. B **96**, 104410 (2017).

[50] S. K. Takada, Y. K. Wakabayashi, Y. Krockenberger, S. Ohya, M. Tanaka, Y. Taniyasu, and H. Yamamoto, Thickness-dependent quantum transport of Weyl fermions in ultra-high-quality $SrRuO_3$ films, Appl. Phys. Lett. **118**, 092408 (2021).

[51] Y. K. Wakabayashi, S. K. Takada, Y. Krockenberger, K. Takiguchi, S. Ohya, M. Tanaka, Y. Taniyasu, and H. Yamamoto, Structural and transport properties of highly Ru-deficient $SrRu_{0.7}O_3$ thin films prepared by molecular beam epitaxy: Comparison with stoichiometric $SrRuO_3$, AIP advances **11**, 035226 (2021).

[52] Y. Senba, H. Ohashi, Y. Kotani, T. Nakamura, T. Muro, T. Ohkochi, N. Tsuji, H. Kishimoto, T. Miura, M. Tanaka, M. Higashiyama, S. Takahashi, Y. Ishizawa, T. Matsushita, Y. Furukawa, T. Ohata, N. Nariyama, K. Takeshita, T. Kinoshita, A. Fujiwara, M. Takata, and S. Goto, Upgrade of beamline BL25SU for soft x-ray imaging and spectroscopy of solid using nano- and micro-focused beams at SPring-8, AIP Conf. Proc. **1741**, 030044 (2016).

[53] T. Nakamura, T. Muro, F. Z. Guo, T. Matsushita, T. Wakita, T. Hirono, Y. Takeuchi, K. Kobayashi, Development of a soft X-ray magnetic circular dichroism spectrometer using a 1.9 T electromagnet at BL25SU of SPring-8, J. Electron Spectrosc. Relat. Phenom. **144-147**, 1035 (2005).

[54] See Supplemental Material at [URL] for details of the samples, growth conditions, experimental set up for XAS and XMCD measurements, Details of the peak assignments of the Ru $M_{2,3}$-edge XAS and XMCD spectra, XAS and XMCD spectra for the SRO films at $\theta$ = 90° and 20°, XMCD sum rules for estimating $m_{orb}/m_{spin}$, deconvolution of the XAS spectra to subtract signals from the substrate, and XMCD-$\mu_0H$ curves.

[55] J. Okamoto, T. Okane, Y. Saitoh, K. Terai, S.-I. Fujimori, Y. Muramatsu, K. Yoshii,1 K. Mamiya, T. Koide, A. Fujimori, Z. Fang, Y. Takeda, and M. Takano, Soft x-ray magnetic circular dichroism study of $Ca_{1-x}Sr_xRuO_3$ across the ferromagnetic quantum phase transition, Phys. Rev. B **76**, 184441 (2007).

[56] K. Terai, K. Yoshii, Y. Takeda, S. I. Fujimori, Y. Saitoh, K. Ohwada, T. Inami, T. Okane, M. Arita, K. Shimada, H. Namatame, M. Taniguchi, K. Kobayashi, M. Kobayashi, and A. Fujimori, X-ray magnetic circular dichroism and photoemission studies of ferromagnetism in $CaMn_{1-x}Ru_xO_3$ thin films, Phys. Rev. B **77**, 115128 (2008).

[57] K. Fujioka, J. Okamoto, T. Mizokawa, A. Fujimori, I. Hase, M. Abbate, H. J. Lin, C. T. Chen, Y. Takeda, and M. Takano, Electronic structure of $SrRuO_3$, Phys. Rev. B **56**, 6380 (1997).

[58] M. Takizawa, D. Toyota, H. Wadati, A. Chikamatsu, H. Kumigashira, A. Fujimori, M. Oshima, Z. Fang, M. Lippmaa, M. Kawasaki, and H. Koinuma, Manifestation of correlation effects in the photoemission spectra of $Ca_{1-x}Sr_xRuO_3$, Phys. Rev. B **72**,

060404(R) (2005).

[59] E. B. Guedes, M. Abbate, K. Ishigami, A. Fujimori, K. Yoshimatsu, H. Kumigashira, M. Oshima, F. C. Vicentin, P. T. Fonseca, and R. J. O. Mossanek, Core level and valence band spectroscopy of SrRuO$_3$: Electron correlation and covalence effects, Phys. Rev. B **86**, 235127 (2012).

[60] N. Nakajima, M. Deguchi, H. Maruyama, K. Ishiji, and Y. Tezuka, X-Ray Spectroscopic Study on Photoluminescence Properties of Red Phosphor SrTiO$_3$:Pr$^{3+}$,Al, Jpn. J. Appl. Phys. **49**, 09ME04 (2010).

[61] W. Siemons, G. Koster, A. Vailionis, H. Yamamoto, D. H. A. Blank, and M. R. Beasley, Dependence of the electronic structure of SrRuO$_3$ and its degree of correlation on cation off-stoichiometry, Phys. Rev. B **76**, 075126 (2007).

[62] M. Gu, J. Laverock, B. Chen, K. E. Smith, S. A. Wolf, and J. Lu, Metal-insulator transition induced in CaVO$_3$ thin films. J. Appl. Phys. **113**, 133704 (2013).

[63] M. Kobayashi, K. Yoshimatsu, E. Sakai, M. Kitamura, K. Horiba, A. Fujimori, and H. Kumigashira, Origin of the Anomalous Mass Renormalization in Metallic Quantum Well States of Strongly Correlated Oxide SrVO$_3$, Phys. Rev. Lett. **115**, 076801 (2015).

[64] D. J. Singh, Electronic and magnetic properties of the 4$d$ itinerant ferromagnet SrRuO3, J. Appl. Phys. **79**, 4818 (1996).

[65] K. Yamagami, Y. Fujisawa, B. Driesen, C. H. Hsu, K. Kawaguchi, H. Tanaka, T. Kondo, Y. Zhang, H. Wadati, K. Araki, T. Takeda, Y. Takeda, T. Muro, F. C. Chuang, Y. Niimi, K. Kuroda, M. Kobayashi, and Y. Okada, Itinerant ferromagnetism mediated by giant spin polarization of the metallic ligand band in the van der Waals magnet Fe$_5$GeTe$_2$, Phys. Rev. B **103**, L060403 (2021).

[66] A. J. Hauser, J. M. Lucy, M. W. Gaultois, M. R. Ball, J. R. Soliz, Y. Choi, O. D. Restrepo, W. Windl, J. W. Freeland, D. Haskel, P. M. Woodward, and F. Yang, Magnetic structure in epitaxially strained Sr$_2$CrReO$_6$ thin films by element-specific XAS and XMCD, Phys. Rev. B **89**, 180402(R) (2014).

**Figures and figure captions**

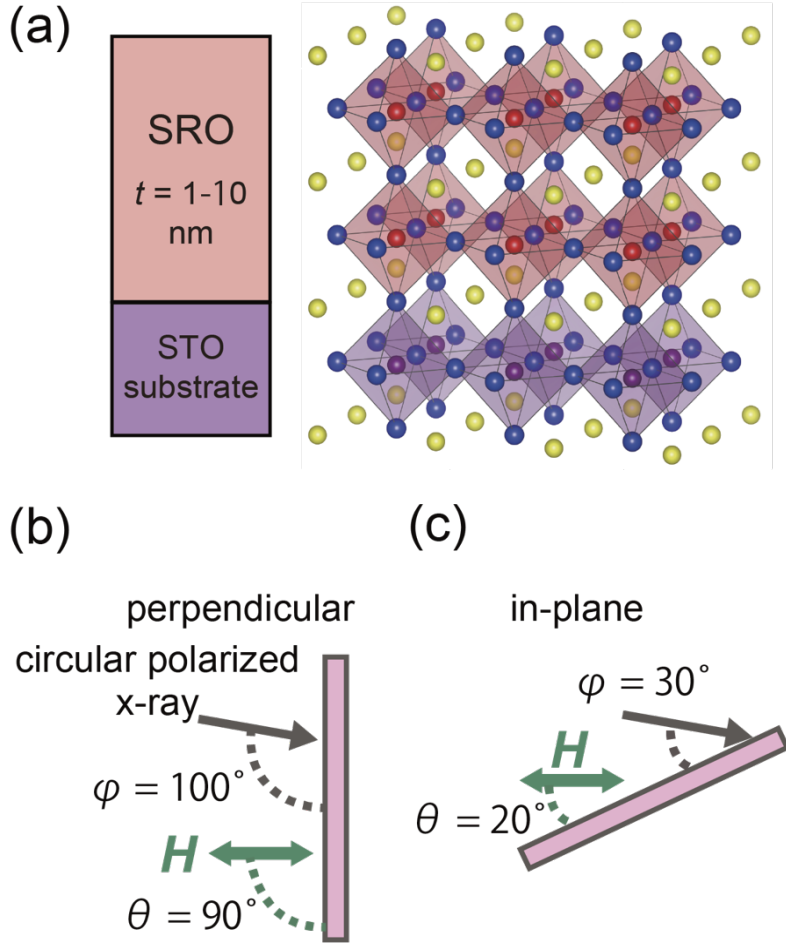

FIG. 1. Schematic diagrams of the sample and crystal structures of the SRO films. In the schematic crystal image, yellow, blue, red, and purple spheres indicate strontium, oxygen, ruthenium, and titanium, respectively. Illustrations of the measurement configurations for the (b) perpendicular and (c) in-plane XAS and XMCD measurements.

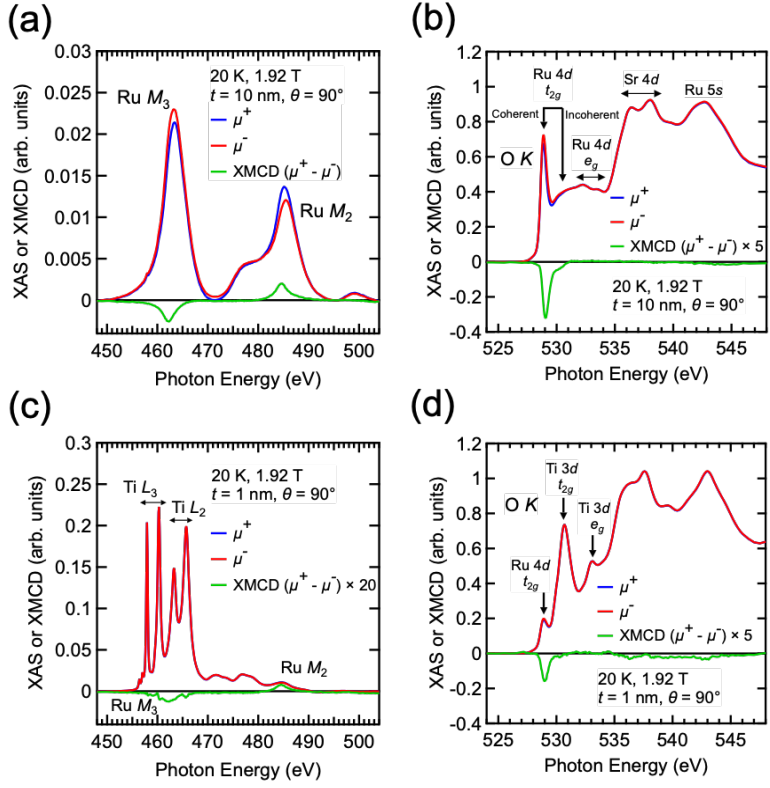

FIG. 2. Ru $M_{2,3}$-edge XAS and XMCD spectra [(a),(c)] and O $K$-edge XAS and XMCD spectra [(b),(d)] for the SRO films with $t$ = 10 nm [(a),(b)] and 1 nm [(c),(d)] at 20 K under a magnetic field $\mu_0 H$ of 1.92 T at $\theta$ = 90°.

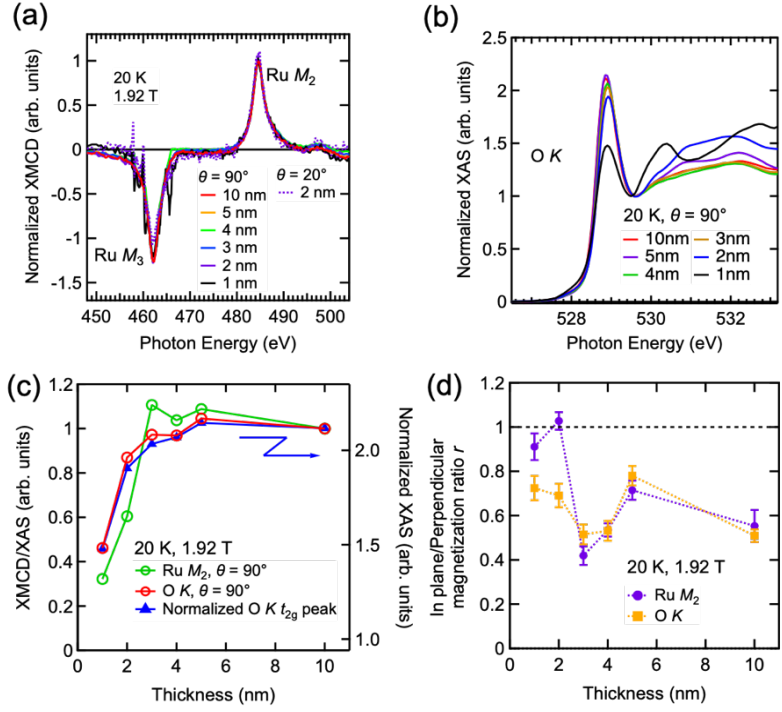

FIG. 3. (a) Ru $M_{2,3}$-edge XMCD spectra normalized at the Ru $M_2$-edge XMCD peak for the SRO films with $t$ = 1-10 nm at $\theta$ = 90° or $\theta$ = 20°. (b) O $K$-edge XAS spectra normalized at 529.5 eV for the SRO films with $t$ = 1-10 nm at $\theta$ = 90° after subtracting the absorption from the STO substrate. (c) Thickness dependence of the Ru $4d$ $t_{2g}$ peak in the O $K$-edge XAS spectrum normalized at 529.5 eV, ratio of the XMCD intensity to the XAS intensity of the Ru $M_2$ edge peak at $\theta$ = 90°, and ratio of the XMCD intensity to the XAS intensity of the O $K$-edge Ru $4d$ $t_{2g}$ peak at $\theta$ = 90°. (d) The ratios of the XMCD/XAS intensity at $\theta$ = 20° to that at $\theta$ = 90°. All measurements were performed at 20 K under a magnetic field $\mu_0 H$ of 1.92 T.

# Supplemental Material for
# Magnetic anisotropy driven by ligand in 4*d* transition-metal oxide SrRuO₃

Yuki K. Wakabayashi,[1,*] Masaki Kobayashi,[2,3] Yuichi Seki,[3] Yoshinori Kotani,[4] Takuo Ohkochi,[4] Kohei Yamagami,[4] Miho Kitamura,[5] Yoshitaka Taniyasu,[1] Yoshiharu Krockenberger,[1] and Hideki Yamamoto[1]

[1]*NTT Basic Research Laboratories, NTT Corporation, Atsugi, Kanagawa 243-0198, Japan*
[2]*Center for Spintronics Research Network, The University of Tokyo, 7-3-1 Hongo, Bunkyo-ku, Tokyo 113-8656, Japan*
[3]*Department of Electrical Engineering and Information Systems, The University of Tokyo, Bunkyo, Tokyo 113-8656, Japan*
[4]*Japan Synchrotron Radiation Research Institute (JASRI), 1-1-1 Kouto, Sayo, Hyogo, 679-5198, Japan*
[5]*Institute for Advanced Synchrotron Light Source, National Institutes for Quantum Science and Technology (QST), Sendai 980-8579, Japan*

[*]Corresponding author: yuuki.wakabayashi@ntt.com

**I. Samples, growth conditions, and experimental set up for XAS and XMCD measurements**

  Epitaxial growth of the SRO films was confirmed by the sharp streaky reflection high energy electron diffraction patterns [49]. The growth parameters were optimized by Bayesian optimization, a machine learning technique [44]. All SRO films were prepared under the same growth conditions as in Refs. [50,51,67], in which the thickness-dependent magnetic and electrical properties together with crystallographic analyses of the SRO films were also reported. The growth rate of 1.05 Å/s was deduced from the thickness calibration of the film using cross-sectional scanning transmission electron microscopy. Further information about the MBE setup and preparation of the substrates is available elsewhere [24]. The residual resistivity ratio (RRR) of the thickest SRO film ($t$ = 10 nm) was 21. This was high enough to observe the intrinsic quantum transport of Weyl fermions in SRO [50], such as quantum oscillations of the high mobility carriers from the bulk three-dimensional Weyl fermions [31] and two-dimensional Weyl fermions from surface Fermi arcs [32], and it is a hallmark of the stoichiometric composition of the films. The availability of such stoichiometric SRO films with precise thicknesses in the range of ≤ 10 nm allows for systematic investigation of the coherent Ru $4d$ $t_{2g}$-O $2p$ hybridized states and magnetic anisotropy of Ru and O ions while the degree of hybridization is being varied.

  XAS and XMCD are the tools of choice for investigating the electronic structures and element-specific magnetic properties of magnetic materials [68,69]. The monochromator resolution $E/\Delta E$ was over 10,000. The designed beam spot size was 10 × 100-200 μm$^2$. For the XMCD measurements, absorption spectra for circularly polarized X rays with the photon helicity parallel ($\mu^+$) or antiparallel ($\mu^-$) to the spin polarization were obtained by reversing the photon helicity at each photon energy $h\nu$ and recording them in the total-electron-yield (TEY) mode. The $\mu^+$ and $\mu^-$ spectra at the Ru $M_{2,3}$ edges and O $K$ edge were taken under both positive and negative applied magnetic fields and averaged to eliminate spurious dichroic signals. In order to estimate the integrated values of the XAS spectra at the Ru $M_{2,3}$ edges, hyperbolic tangent functions as background were subtracted from the spectra.

## II. Details of the peak assignments of the Ru $M_{2,3}$-edge XAS and XMCD spectra for the SRO film with $t$ = 10 nm

Figure 2(a) in the main text shows the Ru $M_{2,3}$-edge XAS and XMCD spectra for the SRO film with $t$ = 10 nm. The absorption peaks from the Ru $3p_{3/2}$ and $3p_{1/2}$ core levels into the Ru $4d$ states appeared at around 463 and 485 eV, respectively, both in the XAS and XMCD spectra [55,56]. The absorption peaks at 477.5 eV (Ru $3p_{3/2}$ → Ru $5s$) and 499 eV (Ru $3p_{1/2}$ → Ru $5s$) which were observed in only XAS can be assigned to the Ru $5s$ states having no spin polarization [55,56]. The Ru $M_3$-edge XMCD peaks are observed at lower energies than the corresponding Ru $M_3$-edge XAS peaks. While the Ru $M_3$-edge XAS and XMCD peaks arise from transitions to unoccupied Ru $4d$ $t_{2g}$ and $e_g$ states [55], the different peak positions between the XAS and XMCD peaks can be explained assuming that only the Ru $4d$ $t_{2g}$ states crossing the Fermi level ($E_F$) exhibit spin polarization. This interpretation is in accordance with our previous Ru $M_{2,3}$ XMCD studies [13] and density functional theory (DFT) calculations [31], which show that the half-metallic Ru $4d$ $t_{2g}$ states cross $E_F$.

## III. Ru $M_{2,3}$-edge and O $K$-edge XAS and XMCD spectra for the SRO films with $t$ = 1-10 nm at $\theta$ = 90° and $\theta$ = 20°

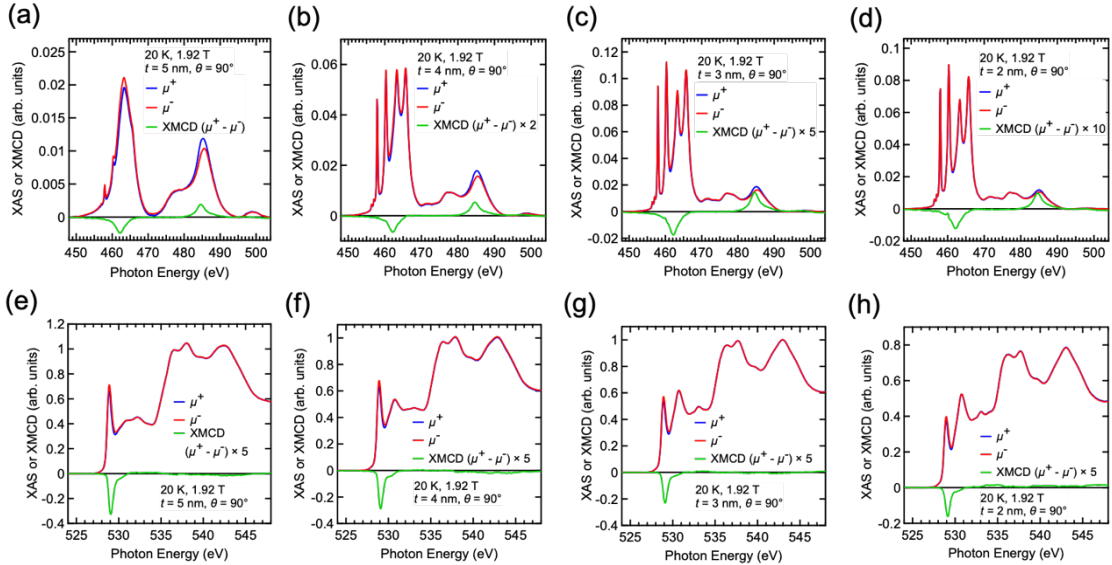

FIG. S1. (a)-(d) Ru $M_{2,3}$-edge XAS and XMCD spectra for the SRO films with $t$ = (a) 5, (b) 4, (c) 3, and (d) 2 nm at 20 K under a magnetic field $\mu_0 H$ of 1.92 T at $\theta$ = 90°. (e)-(h) O $K$-edge XAS and XMCD spectra for the SRO films with $t$ = (e) 5, (f) 4, (g) 3, and (h) 2 nm.

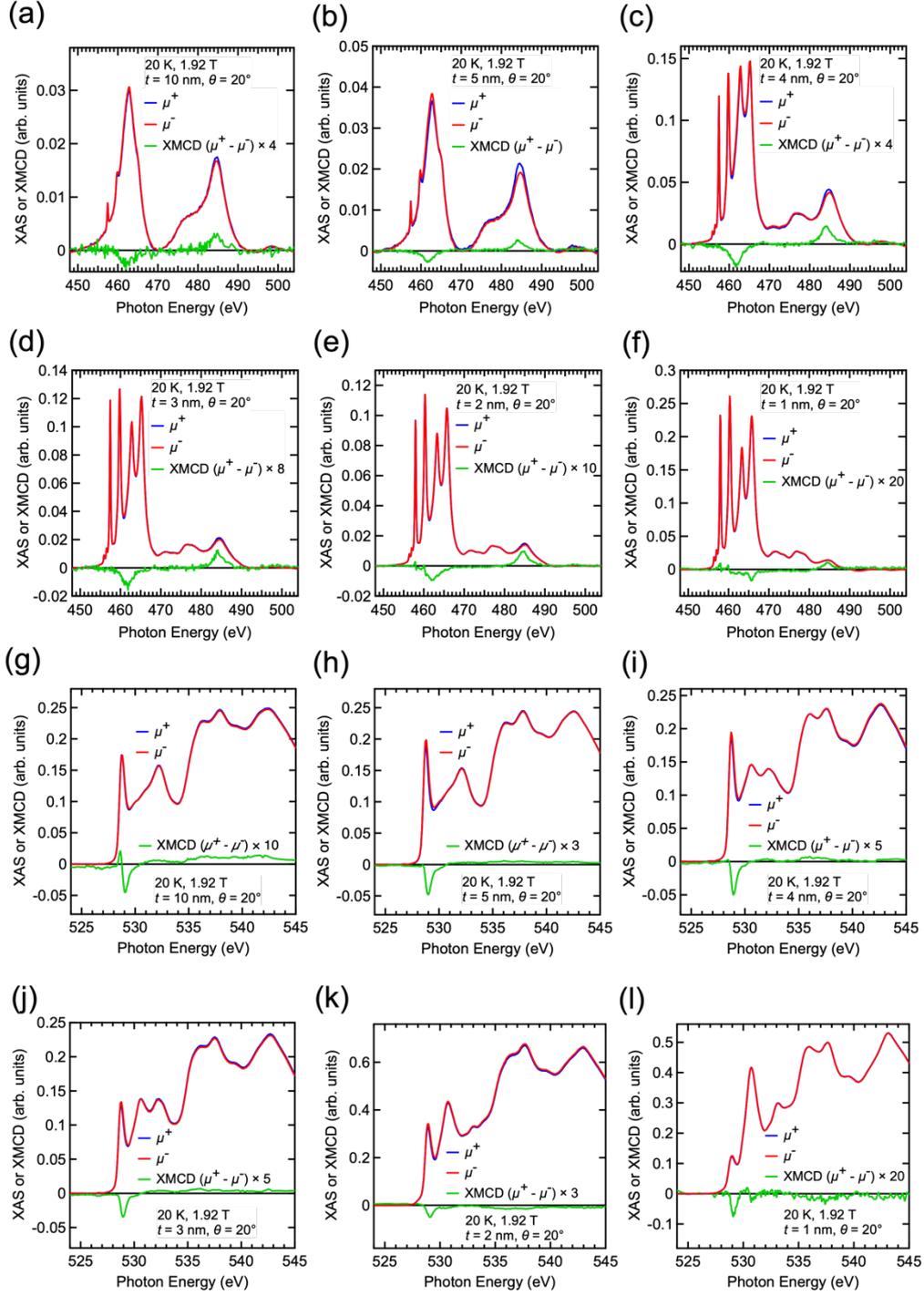

FIG. S2. (a)-(f) Ru $M_{2,3}$-edge XAS and XMCD spectra for the SRO films with $t =$ (a) 10, (b) 5, (c) 4, (d) 3, (e) 2 and (f) 1 nm at 20 K under a magnetic field $\mu_0 H$ of 1.92 T at $\theta = 20°$. (g)-(l) O $K$-edge XAS and XMCD spectra for the SRO films with $t =$ (g) 10, (h) 5, (i) 4, (j) 3, (k) 2 and (l) 1 nm.

## IV. XMCD sum rules for estimating $m_{orb}/m_{spin}$

According to the XMCD sum rules, the orbital magnetic moment $m_{orb}$ and the spin magnetic moment $m_{spin}$ of the Ru$^{4+}$ 4d states are described as follows [46-48]:

$$m_{orb} = -\frac{4(10-n_{4d})}{3r} \int_{M_2+M_3} (\mu^+ - \mu^-) dE, \quad (S1)$$

$$m_{spin} + m_T = -\frac{2(10-n_{4d})}{r} [\int_{M_3} (\mu^+ - \mu^-) dE - 2\int_{M_2} (\mu^+ - \mu^-) dE]. \quad (S2)$$

Here, $r = \int_{M_2+M_3} (\mu^+ + \mu^-) dE$, and $n_{4d}$ is the number of electrons in 4d orbitals, which is assumed to be four. For ions in octahedral symmetry, the magnetic dipole moment $m_T$ is a small number and can be neglected compared to $m_{spin}$ [70]. By dividing equation (S1) by equation (S2), $m_{orb}/m_{spin}$ is expressed by

$$\frac{m_{orb}}{m_{spin}} = \frac{3}{2} \frac{\int_{M_3} (\mu^+ - \mu^-) dE - 2\int_{M_2} (\mu^+ - \mu^-) dE}{\int_{M_2+M_3} (\mu^+ - \mu^-) dE}. \quad (S3)$$

Thus, we can obtain $m_{orb}/m_{spin}$ directly from the XMCD spectra without an assumption of the $n_{4d}$ value.

## V. Deconvolution of the XAS spectra to subtract signals from the substrate

Figures S3(a) and S3(b) show examples of the deconvolution procedure of the XAS spectra for the SRO film with $t = 3$ nm at $\theta = 90°$. The procedure consists of two steps. First, to estimate the contribution of the STO substrate to the XAS spectrum, the raw XAS spectrum (red curves) is fitted by a weighted sum of the XAS spectrum for a pristine STO substrate (black dashed curves) and that for the SRO film with $t = 10$ nm, free from the substrate signals (blue dashed curves). Then, subtraction of the substrate spectrum in accordance with the weighting provides the XAS spectrum inherent to the 3-nm-thick SRO film (black curves). The examples of the deconvolution procedure for the SRO film with $t = 2$ nm at $\theta = 20°$ are also shown in Figs. S3(c) and S3(d).

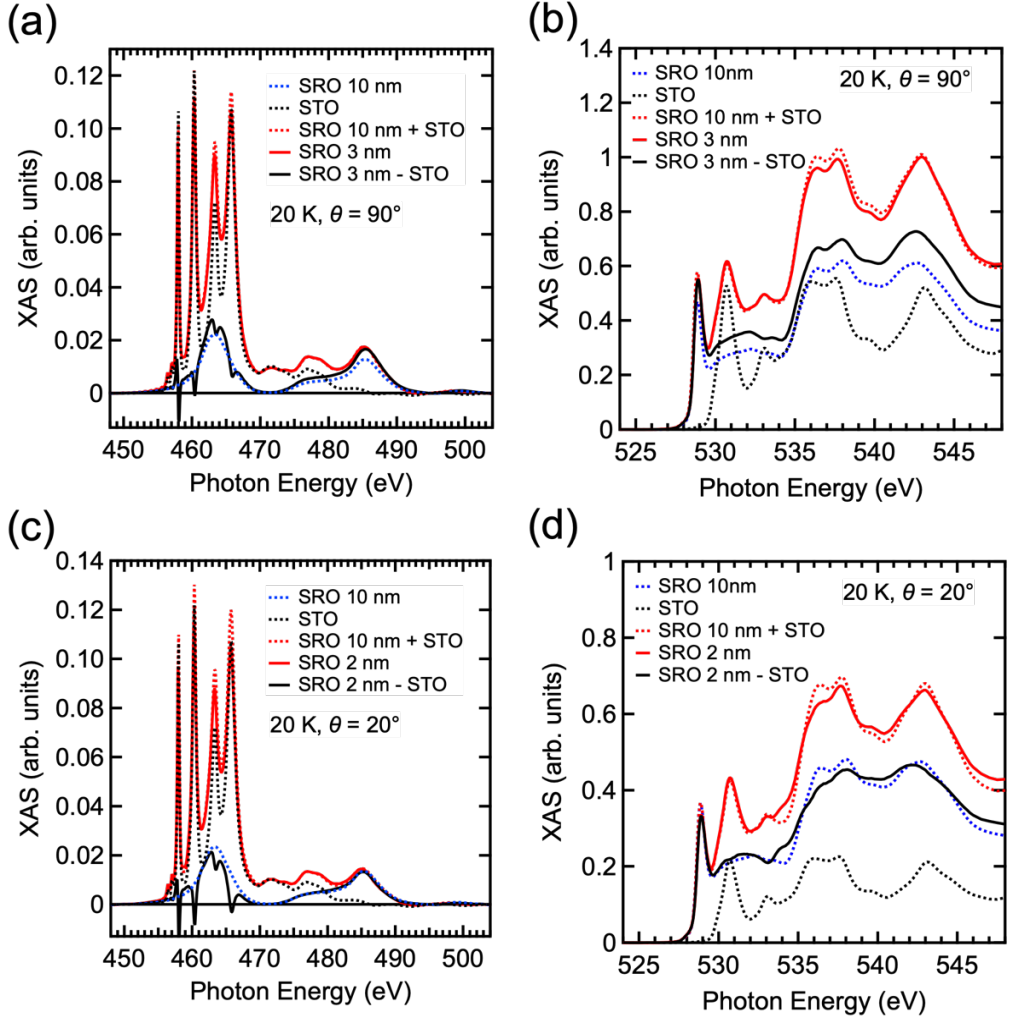

FIG. S3. Ru $M_{2,3}$-edge XAS spectra for the SRO films with $t = 10$ nm (blue dashed curves) and 3 nm (red curve) and for the STO substrate (black dashed curve) at 20 K under a magnetic field $\mu_0 H = 1.92$ T at $\theta = 90°$. (b) O $K$-edge XAS spectra for the SRO films with $t = 10$ nm (blue dashed curve) and 3 nm (red curve) and for the STO substrate (black dashed curve) at 20 K under a magnetic field $\mu_0 H$ of 1.92 T at $\theta = 90°$. The red dashed curves in (a) and (b) are the weighted sum of the XAS spectra for the SRO film with $t = 10$ nm and for the STO substrate. The black curves are the XAS spectra for the SRO film with $t = 3$ nm after subtracting the absorption from the STO substrate. (c),(d) The same deconvolution procedure as that in (a) and (b) for the SRO film with $t = 2$ nm at $\theta = 20°$.

## VI. XMCD-$\mu_0 H$ curves measured at the Ru $M_3$-edge for the SRO film with $t = 1$-10 nm at 20 K with $\theta = 90°$

Figure S4 shows the XMCD-$\mu_0 H$ curves measured at the Ru $M_3$-edge for the SRO film with $t$ = 1-10 nm at 20 K and $\theta$ = 90°. The coercive field $H_c$ increases with decreasing thickness from 10 nm to 2 nm [Fig. S4]. Since the magnetic domains tend to be pinned by grain boundaries and other defects, the large $H_c$ likely stems from disorder near the interface [71]. Notably, $H_c$ at $t$ = 1 nm, contrary to the above described trend of $H_c$, becomes smaller than those at $t \geq 2$ nm, and the Ru magnetic moments for the SRO films with $t$ = 1 and 2 nm at $B$ = 2 T are smaller than those at $t \geq 3$ nm. The small magnetic moments for the SRO films with $t$ = 1 and 2 nm at $B$ = 2 T imply that the SRO films are no longer single-domain magnetic films due to the incoherence of the hybridized Ru 4$d$ $t_{2g}$-O 2$p$ states from disorder-induced localization near the interface or increase in the effective Coulomb interaction with dimensionality reduction [62,63]. Together with the difference in the magnetic anisotropy of Ru and O ions in the SRO films with $t$ = 1 and 2 nm, these results also confirm that the magnetic states of the SRO films with $t$ = 1 and 2 nm are different.

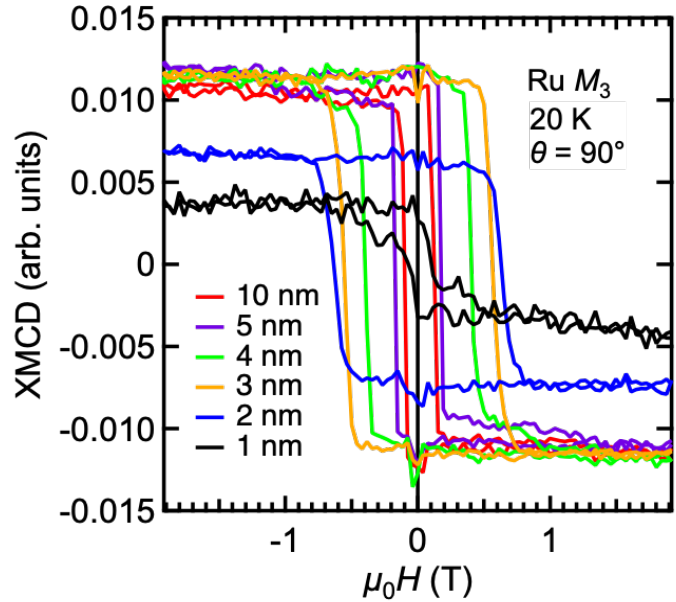

FIG. S4. XMCD-$\mu_0 H$ curves measured at the Ru $M_3$-edge for the SRO film with $t$ = 1-10 nm at 20 K and $\theta$ = 90°.

**REFERENCES**

[67] S. K. Takada, Y. K. Wakabayashi, Y. Krockenberger, H. Irie, S. Ohya, M. Tanaka, Y. Taniyasu, and H. Yamamoto, Scattering-dependent transport of SrRuO$_3$ films: From Weyl fermion transport to hump-like Hall effect anomaly, Phys. Rev. Mater. **7**, 054406 (2023).

[68] Y. K. Wakabayashi, S. Sakamoto, Y. Takeda, K. Ishigami, Y. Takahashi, Y. Saitoh, H. Yamagami, A. Fujimori, M. Tanaka, and S. Ohya, Room-temperature local ferromagnetism and its nanoscale expansion in the ferromagnetic semiconductor Ge$_{1-x}$Fe$_x$, Sci. Rep. **6**, 23295 (2016).

[69] Y. K. Wakabayashi, R. Akiyama, Y. Takeda, M. Horio, G. Shibata, S. Sakamoto, Y. Ban, Y. Saitoh, H. Yamagami, A. Fujimori, M. Tanaka, and S. Ohya, Origin of the large positive magnetoresistance in Ge$_{1-x}$Mn$_x$ granular thin films, Phys. Rev. B **95**, 014417 (2017).

[70] Y. Teramura, A. Tanaka, and T. Jo, J. Phys. Soc. Jpn. **65**, 1053 (1996).

[71] M. Izumi, K. Nakazawa, Y. Bando, Y. Yoneda, and H. Terauchi, Magnetotransport of SrRuO$_3$ Thin Film on SrTiO$_3$ (001), J. Phys. Soc. Jpn. **66**, 3893 (1997).